\documentclass[twocolumn,showpacs,preprintnumbers,amsmath,amssymb]{revtex4}

\usepackage{graphicx}
\usepackage{dcolumn}
\usepackage{bm}

\def\be{\begin{equation}}
\def\ee{\end{equation}}
\def\bea{\begin{eqnarray}}
\def\eea{\end{eqnarray}}
\def\NO{\nonumber}
\def\gev{\mathrm{~GeV}}

\begin{document}


\title{Impact of $\eta_c$ hadroproduction data on charmonium production and polarization within NRQCD framework}


\author{Hong-Fei Zhang$^{1}$}
\author{Zhan Sun$^{2}$}
\author{Wen-Long Sang$^{3}$}
\author{Rong Li$^{4}$}
\affiliation{
$^{1}$ Department of Physics, School of Biomedical Engineering, Third Military Medical University, Chongqing 400038, China. \\
$^{2}$ Department of Physics, Chongqing University, Chongqing 401331, P.R. China \\
$^{3}$ School of Physical Science and Technology, Southwest University, Chongqing 400700, China \\
$^{4}$ Department of Applied Physics, Xi'an Jiaotong University, Xi'an 710049, China
}%
\date{\today}

\begin{abstract}
With the recent LHCb data on $\eta_c$ production and based on heavy quark spin symmetry,
we obtain the long-distance matrix elements for both $\eta_c$ and $J/\psi$ productions,
among which, the color-singlet one for $\eta_c$ is obtained directly by the fit of experiment for the first time.
Using our long-distance matrix elements,
we can provide good description of the $\eta_c$ and $J/\psi$ hadroproduction measurements.
Our predictions on $J/\psi$ polarization are in good agreement with the LHCb data and pass through the two sets of CDF measurements in medium $p_t$ region.
Considering all the possible uncertainties carefully, we obtained quite narrow bands of the $J/\psi$ polarization curves.
\end{abstract}

\maketitle

Nonrelativistic QCD (NRQCD) factorization framework~\cite{Bodwin:1994jh} has gained its reputation from the success in many processes,
among which, heavy quarkonia hadroproduction~\cite{Braaten:1994vv, Cho:1995vh, Cho:1995ce} is one of the most remarkable examples.
Moreover, several groups have accomplished their computer programs for the calculation of QCD corrections to quarkonium related processes.
QCD next-to-leading order (NLO) predictions~\cite{Ma:2010yw, Butenschoen:2011yh, Butenschoen:2010rq, Jia:2014jfa, Bodwin:2014gia, Shao:2014yta}
based on NRQCD achieved good agreement with almost all the experimental measurements on quarkonia hadroproduction.
However, for the $J/\psi$ case, one is still suffering from the ambiguity caused by the freedom in the determination
of the color-octet (CO) long-distance matrix elements (LDMEs)~\cite{Butenschoen:2011yh, Butenschoen:2010rq, Ma:2010yw, Butenschoen:2012px, Chao:2012iv, Gong:2012ug, Shao:2014yta}.
In addition, the $J/\psi$ polarization puzzle is another challenge that NRQCD is facing.
Despite that three groups~\cite{Butenschoen:2012px, Chao:2012iv, Gong:2012ug} have made great efforts to proceed the calculation to NLO in $\alpha_s$,
none of their CO LDMEs can reproduce the recent LHCb data~\cite{Aaij:2013nlm, Aaij:2014qea} with good precision.
On the other hand, many works~\cite{Cho:1995vh, Cho:1995ce, Hao:1999kq, Wang:2014vsa} have proceeded their concerns to the processes
in which no experimental data can be used to extract the LDMEs.
There, they estimate these LDMEs based on heavy quark spin symmetry (HQSS) and velocity scaling rule (VSR).
Nevertheless, the proof of NRQCD factorization does not require the two rules~\cite{Nayak:2005rt, Nayak:2005rw},
hence, the phenomenological test of them is urgent.

Recently, LHCb data~\cite{Aaij:2014bga} on $\eta_c$ produciton came out and provided an opportunity to further investigate these problems.
Ref.~\cite{Biswal:2010xk, Likhoded:2014fta} studied direct $\eta_c$ hadroproduction at leading order (LO) in $\alpha_s$ within NRQCD framework,
however, missing the $^1S_0^{[8]}$ channel.
Since only inclusive and prompt $\eta_c$ production rate has been measured,
one should also consider contributions from $h_c$ feeddown,
the asymptotic behavior of which, in large transverse momentum ($p_t$) limit, scales as $p_t^{-6}$.
According to our previous work~\cite{Wang:2014vsa}, the contribution of this part is negligible comparing with experimental data.
Feeddown contributions from other excited $c\bar{c}$ bound states are even smaller than that from $h_c$,
so that they are not under our consideration.
For direct $\eta_c$ production, up to the order of $v^4$, where $v$ is the typical relative velocity of the constituent quark and antiquark in the quarkonium,
four channels ($^1S_0^{[1]}$, $^1S_0^{[8]}$, $^3S_1^{[8]}$, $^1P_1^{[8]}$) are involved.
Among them, $^3S_1^{[8]}$ channel scales as $p_t^{-4}$ in large $p_t$ limit, while the other three scale as $p_t^{-6}$.
Moreover, the NLO QCD corrections to all the channels are not significant, which indicates good convergence in $\alpha_s$ expansion.
Therefore, it is possible to determine $\langle O^{\eta_c}(^3S_1^{[8]})\rangle$ precisely by the fit of the experimental data.
Further, we can assume HQSS and fix the other two CO LDMEs for $\eta_c$ produciton as well as those for $J/\psi$ production,
and see whether they are able to provide reasonable descriptions of $J/\psi$ production and polarization.
Noticing that, in Ref.~\cite{Butenschoen:2011yh, Chao:2012iv, Gong:2012ug}, the LDMEs obtained by minimizing $\chi^2$ do not indicate the VSR,
we give up employing this rule as the basis of our arguement.

We should also notice that the values of the production LDMEs
$\langle O^{\eta_c}(^1S_0^{[1]})\rangle$ and $\langle O^{J/\psi}(^3S_1^{[1]})\rangle$ have never been obtained directly from the fit of experiment;
only the values of the decay ones have been extracted from experiment.
The production LDMEs are considered to be the same as the decay ones in the sense of VSR,
the importance of the higher order effects of which is not clear.
Since the absolute values of the CS LDMEs play a very important role in the exclusive double charmonia production in $e^+e^-$ collisions,
high-precision determination of them would be urgent.
LHCb data on $\eta_c$ production rate provides an opportunity to obtain the value of $\langle O^{\eta_c}(^1S_0^{[1]})\rangle$ by fitting experimental data.
As a result, precise evaluation of the short-distance coefficient (SDC) of $^1S_0^{[1]}$ channel is necessary,
and our calculation will be accurate to NLO in $\alpha_s$ as well as in $v^2$,
while, higher order corrections are neglected.
We should also consider the uncertainty caused by the possible large logarithmic terms involving $E_{\eta_c}$ (the energy of $\eta_c$),
which is brought in by the large rapidity (denoted as $y$) in the LHCb experimental condition.

Since there are only seven experimental data points on $p_t$ distribution of $\eta_c$ production rate,
it is impossible to determine all the LDMEs without further constrainsts.
So, we base our work on HQSS,
and employ the relations of the LDMEs for direct $J/\psi$ produciton obtained in Ref.~\cite{Ma:2010yw, Shao:2014yta},
\bea
M_0=\langle O^{J/\psi}(^1S_0^{[8]})\rangle+r_0\frac{\langle O^{J/\psi}(^3P_0^{[8]})\rangle}{m_c^2}, \NO \\
M_1=\langle O^{J/\psi}(^3S_1^{[8]})\rangle+r_1\frac{\langle O^{J/\psi}(^3P_0^{[8]})\rangle}{m_c^2}, \label{eqn:m}
\eea
where
\bea
&&M_0=(7.4\pm 1.9)\times 10^{-2}\gev^3,~~~~r_0=3.9, \label{eqn:mr} \\
&&M_1=(0.05\pm 0.02)\times 10^{-2}\gev^3,~~~~r_1=-0.56. \NO
\eea
Notice that the universally used definition of the LDMEs are spin- and color-summed,
the relations between the LDMEs for $\eta_c$ and $J/\psi$ based on HQSS are
\bea
&&\langle O^{\eta_c}(^1S_0^{[n]})\rangle=\frac{1}{3}\langle O^{J/\psi}(^3S_1^{[n]})\rangle, \NO \\
&&\langle O^{\eta_c}(^3S_1^{[8]})\rangle=\langle O^{J/\psi}(^1S_0^{[8]})\rangle, \label{eqn:ss} \\ 
&&\langle O^{\eta_c}(^1P_1^{[8]})\rangle=3\times\langle O^{J/\psi}(^3P_0^{[8]})\rangle, \NO \\
&&\langle O^{h_c}(^1P_1^{[1]}/^1S_0^{[8]})\rangle=\langle O^{\chi_{c1}}(^3P_1^{[1]}/^3S_1^{[8]})\rangle, \NO
\eea
where $n$ denotes 1 or 8, corresponding to color-singlet (CS) or CO, respectively.
To determine the value of $\langle O^{\eta_c}(^3S_1^{[8]})\rangle$,
we translate the relation in Eq.(\ref{eqn:m}) into the $\eta_c$ version as
\bea
\frac{\langle O^{\eta_c}(^1P_1^{[8]})\rangle}{m_c^2}&=&\frac{3}{r_0}(M_0-\langle O^{\eta_c}(^3S_1^{[8]})\rangle), \label{eqn:rldme} \\
\langle O^{\eta_c}(^1S_0^{[n]})\rangle&=&\frac{M_1}{3}-\frac{r_1}{3r_0}(M_0-\langle O^{\eta_c}(^3S_1^{[8]})\rangle). \NO
\eea
Then we obtain the equation for fit,
\bea
&&f_{^1S_0^{[1]}}\langle O^{\eta_c}(^1S_0^{[1]})\rangle+(f_{^3S_1^{[8]}}+\frac{r_1}{3r_0}f_{^1S_0^{[8]}}-\frac{3m_c^2}{r_0}f_{^1P_1^{[8]}}) \NO \\
&&\times\langle O^{\eta_c}(^3S_1^{[8]})\rangle=\sigma_{exp}-\sigma_{h_c} \label{eqn:fit} \\
&&+M_0(\frac{r_1}{3r_0}f_{^1S_0^{[8]}}-\frac{3}{r_0}m_c^2f_{^1P_1^{[8]}})-\frac{M_1}{3}f_{^1S_0^{[8]}}, \NO
\eea
where $\sigma_{exp}$, $\sigma_{h_c}$ and $f_n$ denote the experimental data,
the contribution from $h_c$ feeddown, and the SDC for the state $n$, respectively.
Without Eq.(\ref{eqn:fit}), the CS LDME cannot be determined precisely.
Since, on one hand, the SDCs of $^1S_0^{[8]}$ and $^1P_1^{[8]}$ have the same $p_t$ behavior with the CS one,
only the summation of the LDMEs of the three channels can be fixed.
On the other hand, $M_0$ and $M_1$ also have uncertainties which might affect those of the CO LDMEs to be obtained.
Eq.(\ref{eqn:fit}) seperates the CS SDC from the CO ones,
at the same time, the errors from $M_0$, $M_1$ and $\sigma_{h_c}$ are combined with the experimental ones naturally.

To obtain the SDCs, we employ the FDC package~\cite{Wang:2004du}.
In the numerical calculation, we have the following common choices.
$|R_{h_c}'(0)|^{2}=0.075\gev^5$~\cite{Eichten:1995ch} for both LO and NLO calculation, $m_{c}=1.5\gev$, and $v^2=0.23$.
We employ CTEQ6M~\cite{Pumplin:2002vw} as parton distribution function and two-loop $\alpha_s$ running for up-to-NLO calculation,
and CTEQ6L1~\cite{Pumplin:2002vw} and one-loop $\alpha_s$ running for LO.
The branching ratio~\cite{Agashe:2014kda} of $h_c$ to $\eta_c$ is ${\cal B}(h_c\rightarrow\eta_c\gamma)=(51\pm 6)\%$.
Having got the SDCs, after a short calculation, we find that, in Eq.(\ref{eqn:fit}),
the terms involving $h_c$, $^1S_0^{[8]}$ and $^1P_1^{[8]}$ are negligible (less than 2\% of the dominant terms).
Eventually, Eq.(\ref{eqn:fit}) reduces to
\be
f_{^1S_0^{[1]}}\langle O^{\eta_c}(^1S_0^{[1]})\rangle+f_{^3S_1^{[8]}}\langle O^{\eta_c}(^3S_1^{[8]})\rangle=\sigma_{exp}. \label{eqn:fits}
\ee

Eq.(\ref{eqn:fits}) provides an excellent opportunity to determine both of the LDMEs,
$\langle O^{\eta_c}(^1S_0^{[1]})\rangle$ and $\langle O^{\eta_c}(^3S_1^{[8]})\rangle$.
Firstly, the $p_t$ behaviors of $f_{^1S_0^{[1]}}$ and $f_{^3S_1^{[8]}}$ are different,
which is unlike the $J/\psi$ case where the $^3S_1^{[8]}$ and $^3P_J^{[8]}$ channels are entangled.
Further, higher order terms in $\alpha_s$ expansion of both of the SDCs might not be significant~\cite{Ma:2014svb}.
We can expect NLO results to give reliable predictions.

To fix the values of the LDMEs in Eq.(\ref{eqn:fits}), we should first make sure that our SDCs are evaluated properly.
For the CS channel, we would like to obtain the absolute value of the LDME,
to this end, the corresponding SDC should be evaluted precisely.
Hence, both QCD and relativistic corrections are considered here,
while higher order corrections are dropped.
Since the rapidity for the LHCb experimental condition is large, i.e. $2<y<4.5$,
we also consider the uncertainty coming from possible large logarithmic terms brought in by the large scale, $E_{\eta_c}\approx m_te^y/2$, where $m_t=\sqrt{m_{\eta_c}^2+p_t^2}$.
As a result, we calculate the SDC for $^1S_0^{[1]}$ at both $\mu_R=\mu_F=m_t$ and $\mu_R=\mu_F=E_{\eta_c}$, and investigate the corresponding uncertainty,
where $\mu_R$ and $\mu_F$ denote the renormalization and factorization scales, respectively.
For the $^3S_1^{[8]}$ channel, we should be careful.
Eq.(\ref{eqn:m}) and Eq.(\ref{eqn:mr}) are obtained in the absence of relativistic corrections, where only QCD corrections are considered.
To be consistent, at the same time, noticing that relativistic correction contributes a part proportional to the QCD LO (as well as NLO) SDC,
when $p_t$ is larger than about 7GeV~\cite{Xu:2012am},
we should also give up the relativistic-correction contributions to the $^3S_1^{[8]}$ channel.
For the same reason, we fix $\mu_R$ and $\mu_F$ to be $m_t$ in the calculation of the $^3S_1^{[8]}$ SDC.
The relativstic correction contribution and the difference coming from employing another scale are considered to be absorbed into the corresponding LDME.
Through out the rest of this paper, when refering to CO channels,
we adopt the same scheme.

Now, we fit our theoretical predictions to the LHCb data on $p_t$ distribution of $\eta_c$ production rate at both 7TeV and 8TeV presented in Ref.~\cite{Aaij:2014bga},
and obtain the LDMEs in Eq.(\ref{eqn:fits}).
For $\mu_R=\mu_F=m_t$,
the LDMEs are given as
\bea
\langle O^{\eta_c}(^1S_0^{[1]})\rangle&=&(0.16\pm 0.08)\gev^3, \NO \\
\langle O^{\eta_c}(^3S_1^{[8]})\rangle&=&(0.74\pm 0.30)\times 10^{-2}\gev^3 \label{eqn:ldmemt}
\eea
and the $\chi^2/d.o.f=0.15$.
For $\mu_R=\mu_F=E_{\eta_c}$, they are
\bea
\langle O^{\eta_c}(^1S_0^{[1]})\rangle&=&(0.20\pm 0.10)\gev^3, \NO \\
\langle O^{\eta_c}(^3S_1^{[8]})\rangle&=&(0.86\pm 0.27)\times 10^{-2}\gev^3, \label{eqn:ldmeee}
\eea
and the $\chi^2/d.o.f=0.17$.
We get a relatively large uncertainty of the LDMEs in Eq.(\ref{eqn:ldmemt}) and Eq.(\ref{eqn:ldmeee}),
which is due to the large error of the experimental data.
We simply estimate the possible range of $\langle O^{\eta_c}(^1S_0^{[1]})\rangle$ to be from 0.08 to 0.3$\gev^3$,
which is comparable with the values obtained in most of other existing works.
(e.g. $\langle O^{\eta_c}(^1S_0^{[1]})\rangle=0.39\gev^3$ in Ref.~\cite{Eichten:1995ch} and
$\langle O^{\eta_c}(^1S_0^{[1]})\rangle=0.437^{+0.111}_{-0.105}\gev^3$ in Ref.~\cite{Bodwin:2007fz}).

\begin{figure}
\center{
\includegraphics*[scale=0.24]{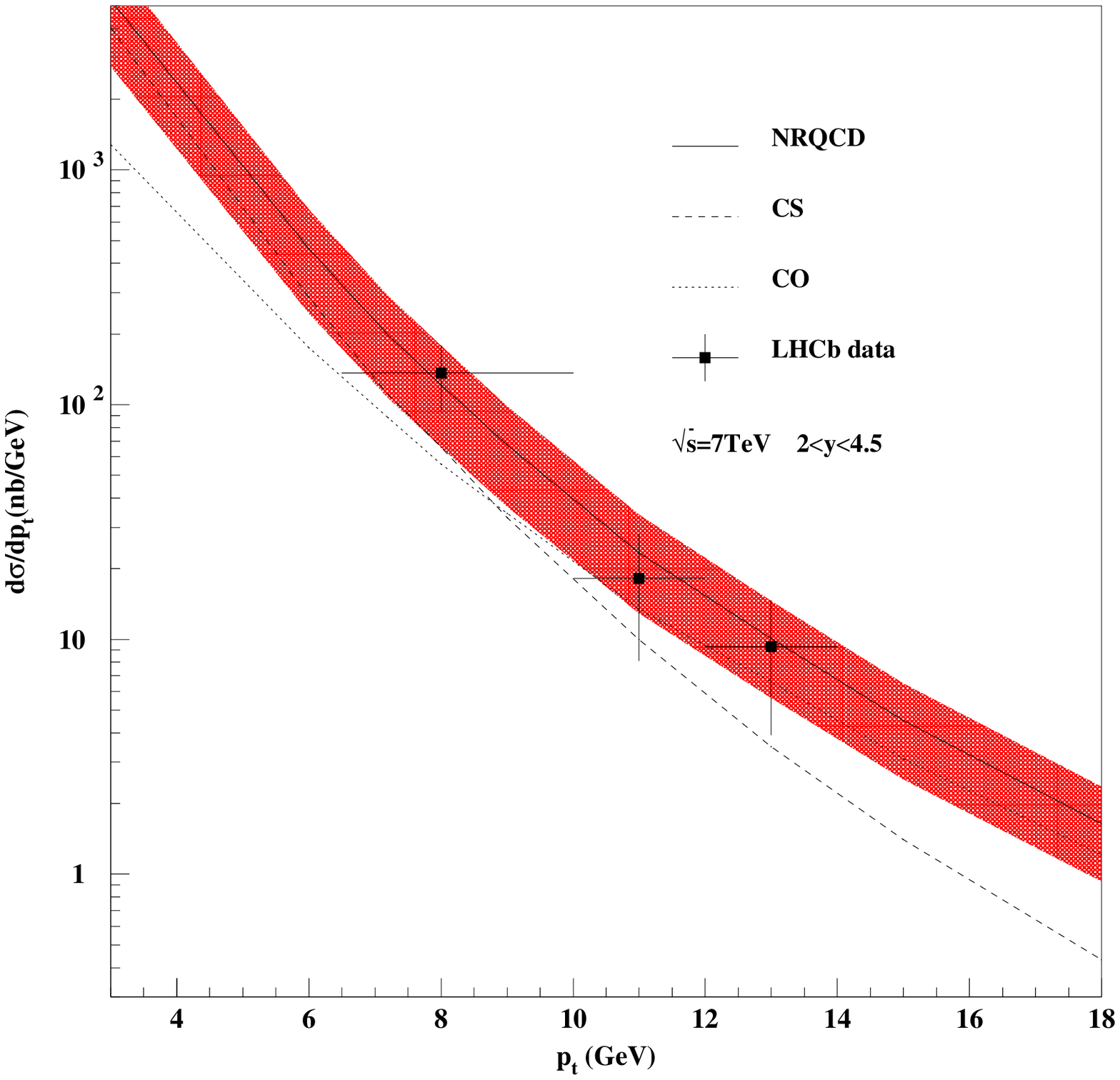}
\includegraphics*[scale=0.24]{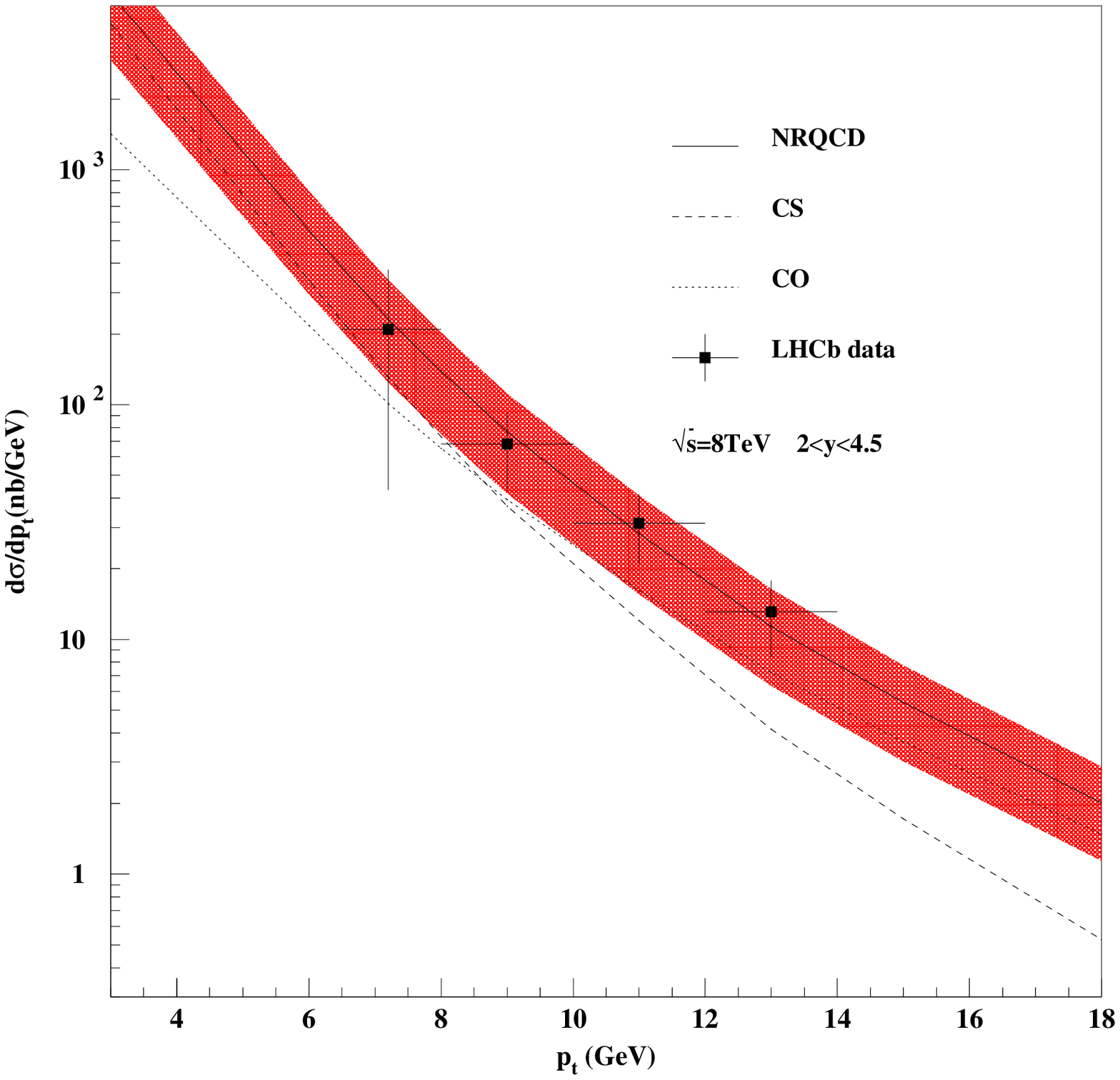}\\
\includegraphics*[scale=0.24]{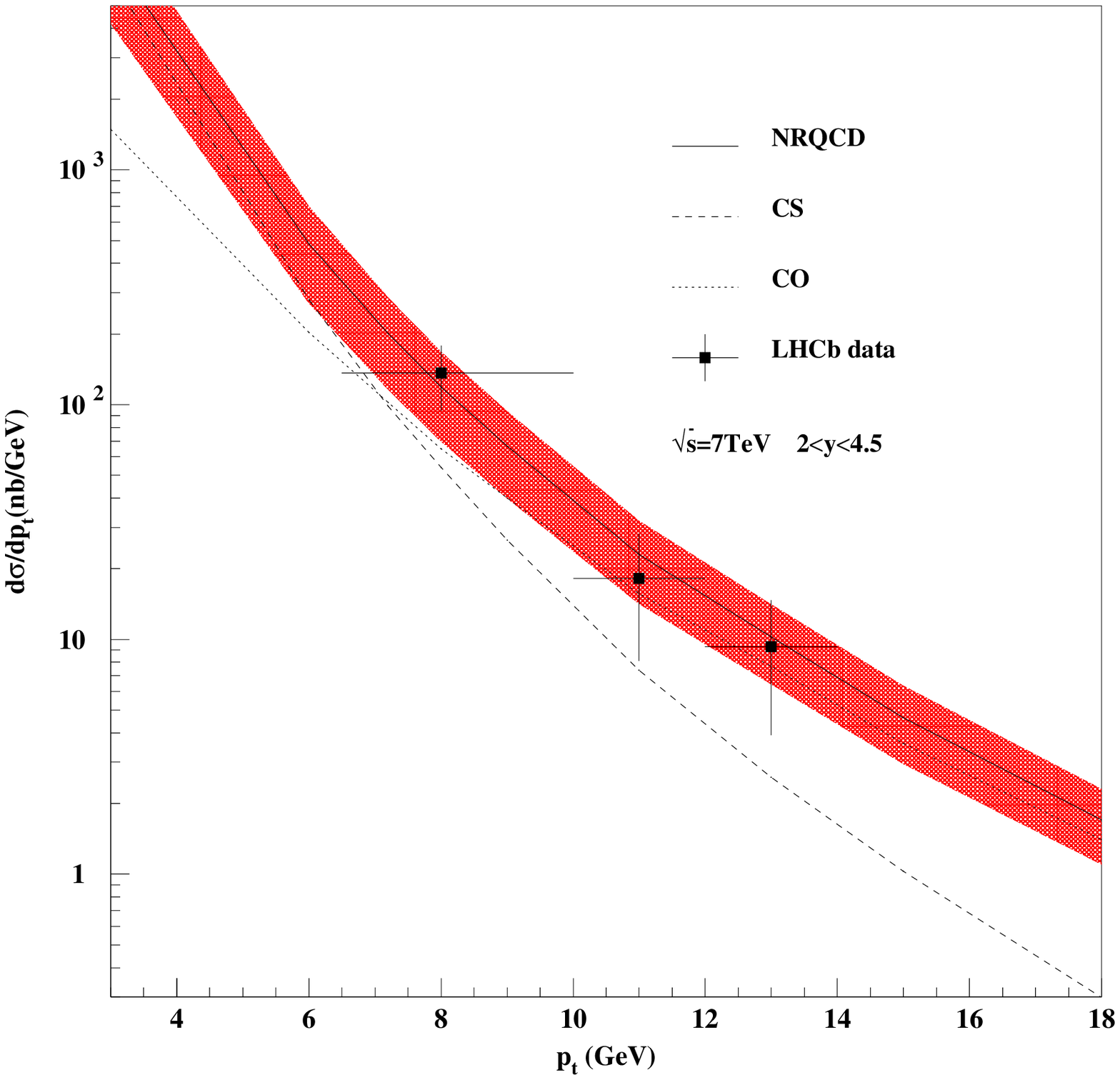}
\includegraphics*[scale=0.24]{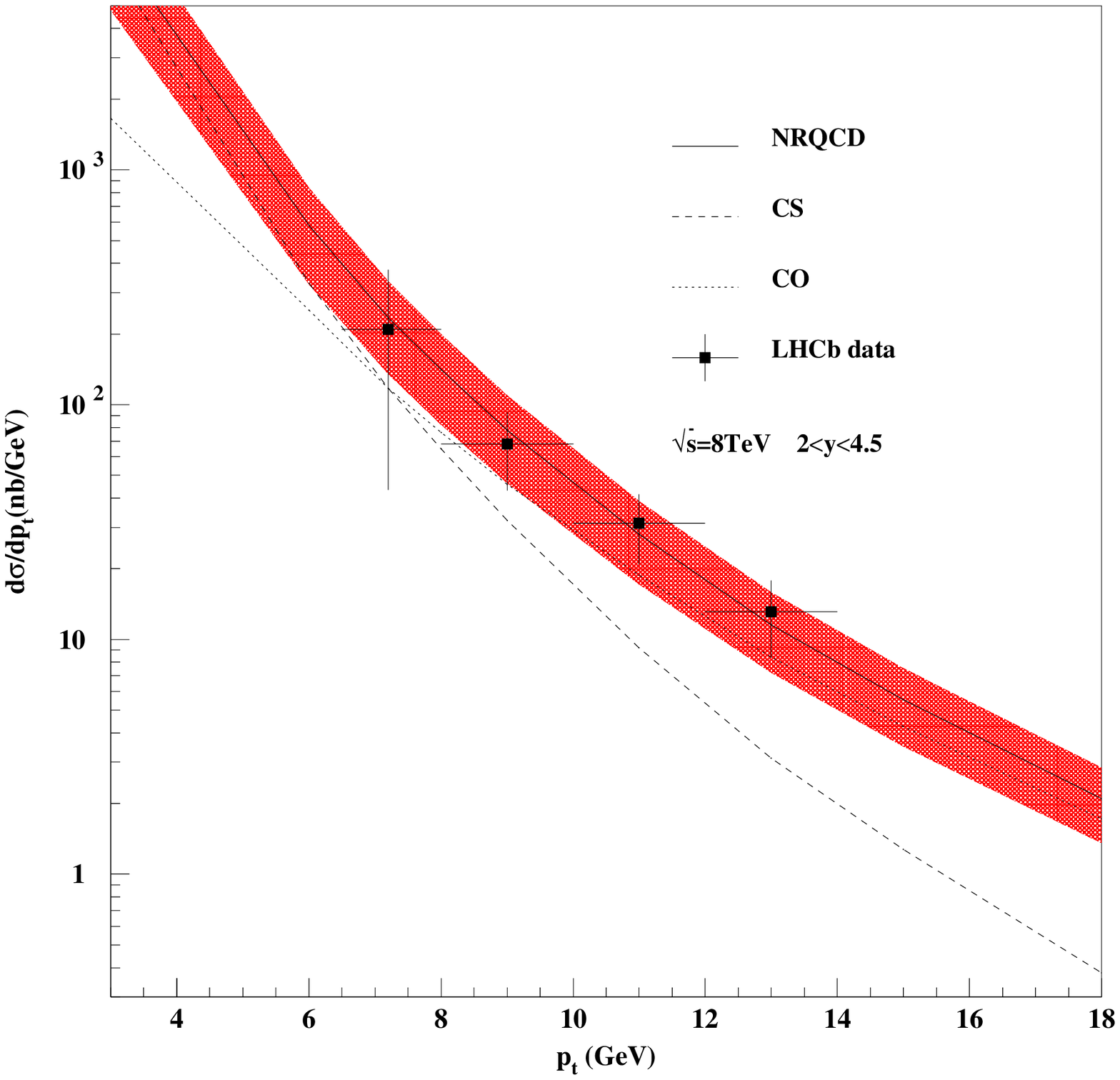}
\caption {\label{fig:etac}
$p_t$ distribution of $\eta_c$ hadroproduction.
The upper and lower plots correspond to $\mu_R=\mu_F=m_t$ and $\mu_R=\mu_F=E_{\eta_c}$, respectively.
The experimental data are taken from Ref.~\cite{Aaij:2014bga}.
}}
\end{figure}

The $p_t$ distribution of $\eta_c$ hadroproduction rate is shown in Fig.\ref{fig:etac}.
We can see that our theoretical prediction can explain the experimental data for both of the choices of the scales.
Also, we can evaluate the integrated cross sections for $\eta_c$ hadroproduction
in the kinematic range $p_t>6.5\gev$ and $2<y<4.5$ at the centre-of-mass energy of 7TeV and 8TeV as
$\mathrm{(\sigma_{\eta_c(1s)})_{\sqrt{s}=7TeV}=(0.53\pm 0.24) \mu b}$ and $\mathrm{(\sigma_{\eta_c(1s)})_{\sqrt{s}=8TeV}=(0.62\pm 0.28) \mu b}$, respectively,
which are consistent with the LHCb measurement~\cite{Aaij:2014bga},
where $\mathrm{(\sigma_{\eta_c(1s)})_{\sqrt{s}=7TeV}=(0.52\pm 0.08\pm 0.09\pm 0.06) \mu b}$ and $\mathrm{(\sigma_{\eta_c(1s)})_{\sqrt{s}=8TeV}=(0.59\pm 0.11\pm 0.09\pm 0.08) \mu b}$, respectively.

Using HQSS and Eq.(\ref{eqn:m}), we can derive the LDMEs for $J/\psi$ production using the second equation in Eq.(\ref{eqn:ss}) and
\bea
\langle O^{J/\psi}(^3S_1^{[8]})\rangle&=&M_1+\frac{r_1}{r_0}\langle O^{J/\psi}(^1S_0^{[8]})\rangle-\frac{r_1}{r_0}M_0, \NO \\
\frac{\langle O^{J/\psi}(^3P_0^{[8]})\rangle}{m_c^2}&=&\frac{M_0}{r_0}-\frac{1}{r_0}\langle O^{J/\psi}(^1S_0^{[8]})\rangle. \label{eqn:ldmef}
\eea
We find that the values of $\langle O^{J/\psi}(^3S_1^{[8]})\rangle$ and $\langle O^{J/\psi}(^3P_0^{[8]})\rangle$ are not sensitive to the values of $\langle O^{\eta_c}(^3S_1^{[8]})\rangle$ and $M_1$.
The major uncerntainty of the two LDMEs comes from the uncertainty of $M_0$.
The large errors in Eq.(\ref{eqn:ldmemt}) and Eq.(\ref{eqn:ldmeee}) only affect the other two CO LDMEs for $J/\psi$ production slightly.
And we obtain
\bea
&&0.24\gev^3<\langle O^{J/\psi}(^3S_1^{[1]})\rangle<0.90\gev^3, \NO \\
&&0.44\times 10^{-2}\gev^3<\langle O^{J/\psi}(^1S_0^{[8]})\rangle<1.13\times 10^{-2}\gev^3, \NO \\
&&\langle O^{J/\psi}(^3S_1^{[8]})\rangle=(1.0\pm 0.3)\times 10^{-2}\gev^3, \label{eqn:LDMEJ} \\
&&\frac{\langle O^{J/\psi}(^3P_0^{[8]})\rangle}{m_c^2}=(1.7\pm 0.5)\times 10^{-2}\gev^3. \NO
\eea
The LDMEs obtained here are consistent with the VSR,
while in most of the existing versions of the CO LDMEs for NLO calculation,
the values of $\langle O^{J/\psi}(^1S_0^{[8]})\rangle$ are one order of magnitude larger than the values of the corresponding $\langle O^{J/\psi}(^3S_1^{[8]})\rangle$.

\begin{figure}
\center{
\includegraphics*[scale=0.24]{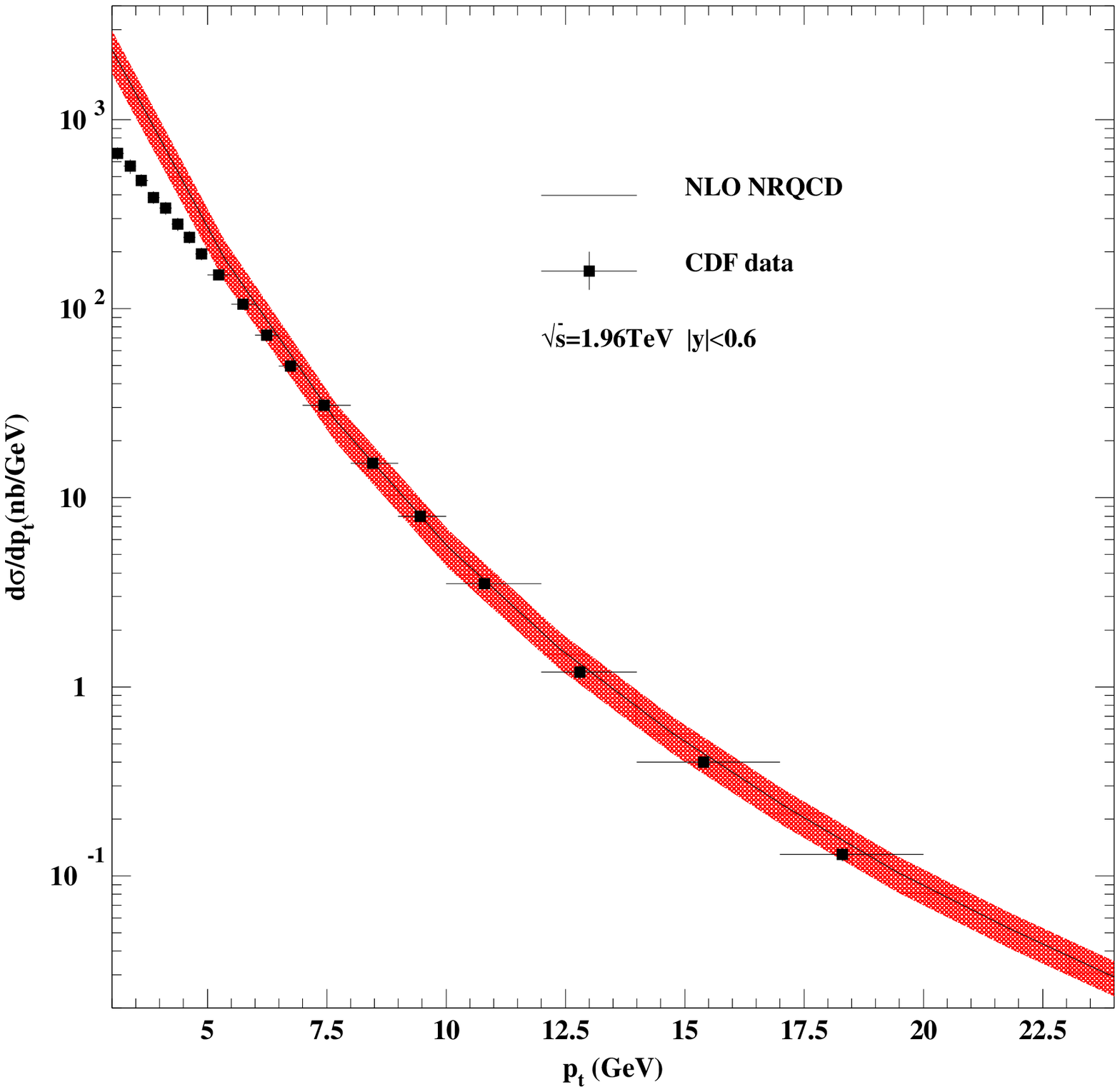}
\includegraphics*[scale=0.24]{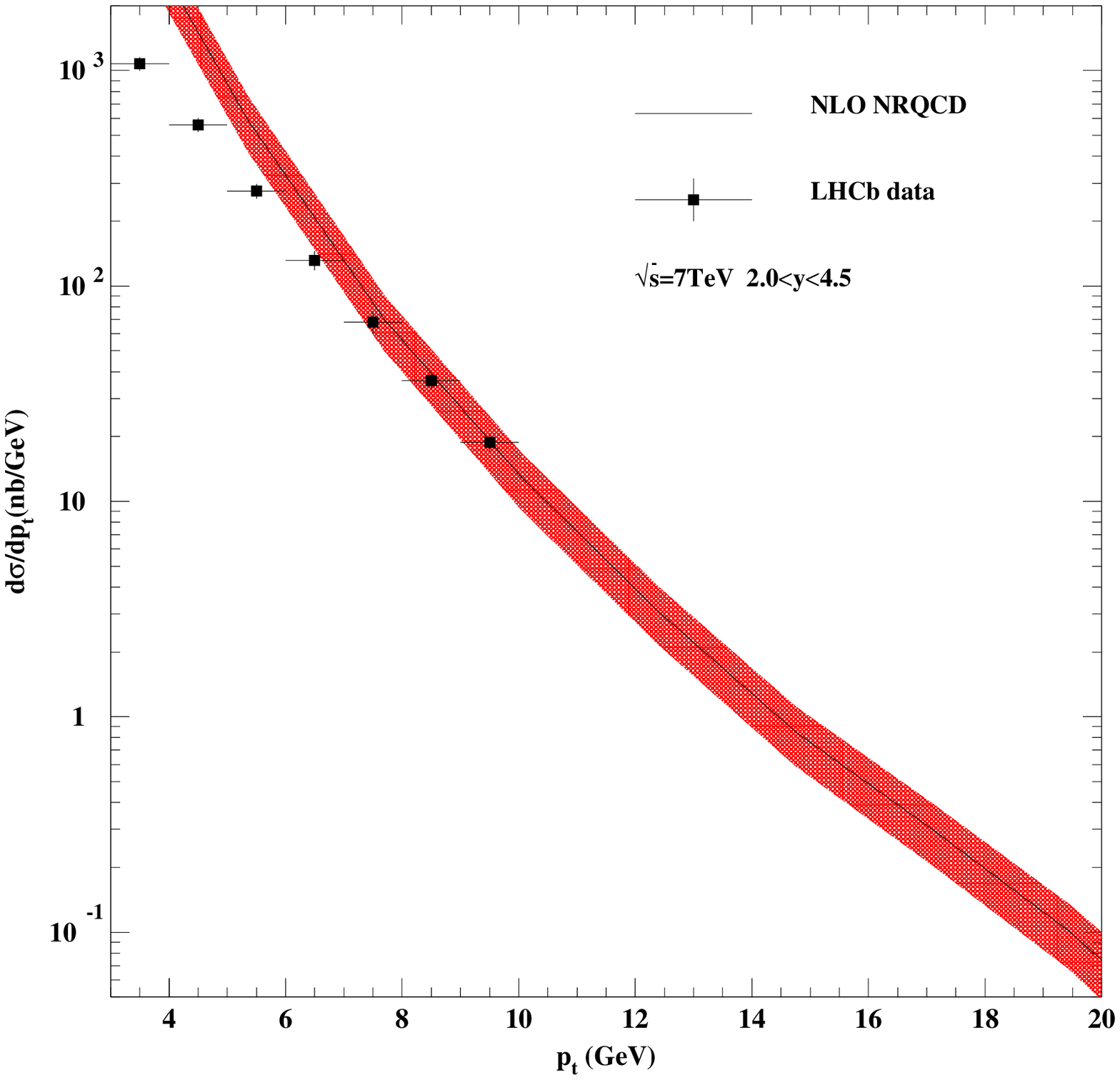}
\caption {\label{fig:pt}
$p_t$ distribution of $J/\psi$ production rate. The CDF and LHCb data are taken from Ref.~\cite{Acosta:2004yw, Aaij:2012ag, LHCb:2012af, Aaij:2011jh}
}}
\end{figure}

Using the LDMEs in Eq.(\ref{eqn:LDMEJ}), we present the results for $J/\psi$ yield in Fig.\ref{fig:pt}.
The LHCb data are obtained by subtracting feeddown contributions of $\psi(2s)$~\cite{Aaij:2012ag} and $\chi_c$~\cite{LHCb:2012af} from the prompt one~\cite{Aaij:2011jh}.
As for the CDF data~\cite{Acosta:2004yw}, lacking measurements on $\chi_c$ feeddown contributions,
we also calculate the production rate of $J/\psi$ coming from $\chi_c$ feeddown based on our previous work~\cite{Gong:2012ug, Jia:2014jfa},
while the $\psi(2s)$ part is ommited~\cite{Aaltonen:2009dm}.
Even though the values of our LDMEs are quite different from those in Ref.~\cite{Butenschoen:2011yh, Chao:2012iv, Gong:2012ug},
they are also able to explain the CDF and LHCb data for $J/\psi$ production well.
This indicates that the three CO SDCs are linear correlated and only two linear combinations of the three CO LDMEs can be fixed stably through the hadroproduction experiment~\cite{Ma:2010yw}.
To present the uncertainty, we should be careful and look at Eq.(\ref{eqn:ldmef}).
$\langle O^{J/\psi}(^3S_1^{[8]})\rangle$ and $\langle O^{J/\psi}(^3P_0^{[8]})\rangle$ should vary their values accordingly
and reach their maximum or minimum at the same time, since all their uncertainties have the same origin, $M_0$.
The bands presented in Fig.\ref{fig:pt} come from the uncertainties of the LDMEs in this sense.

\begin{figure}
\center{
\includegraphics*[scale=0.4]{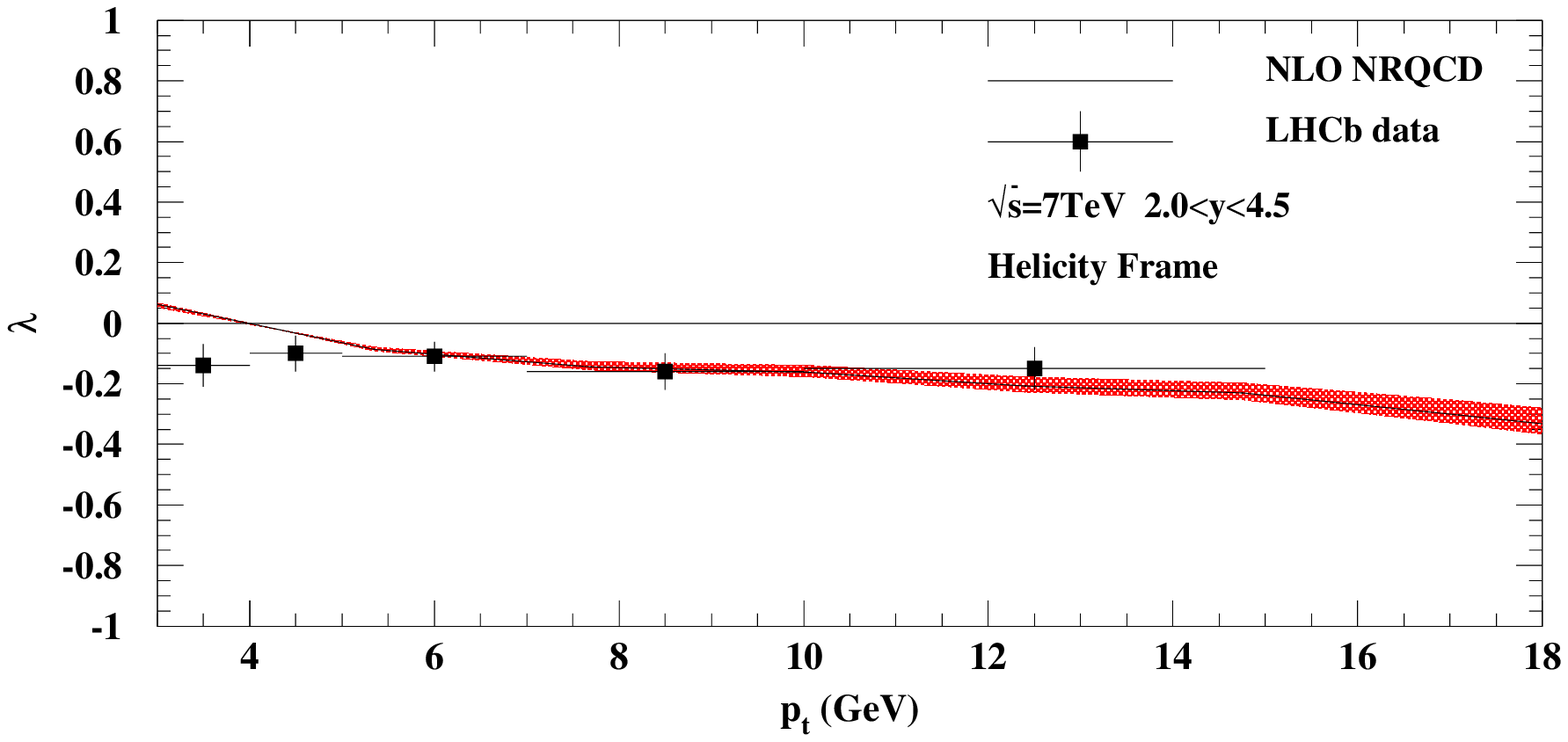}\\
\includegraphics*[scale=0.4]{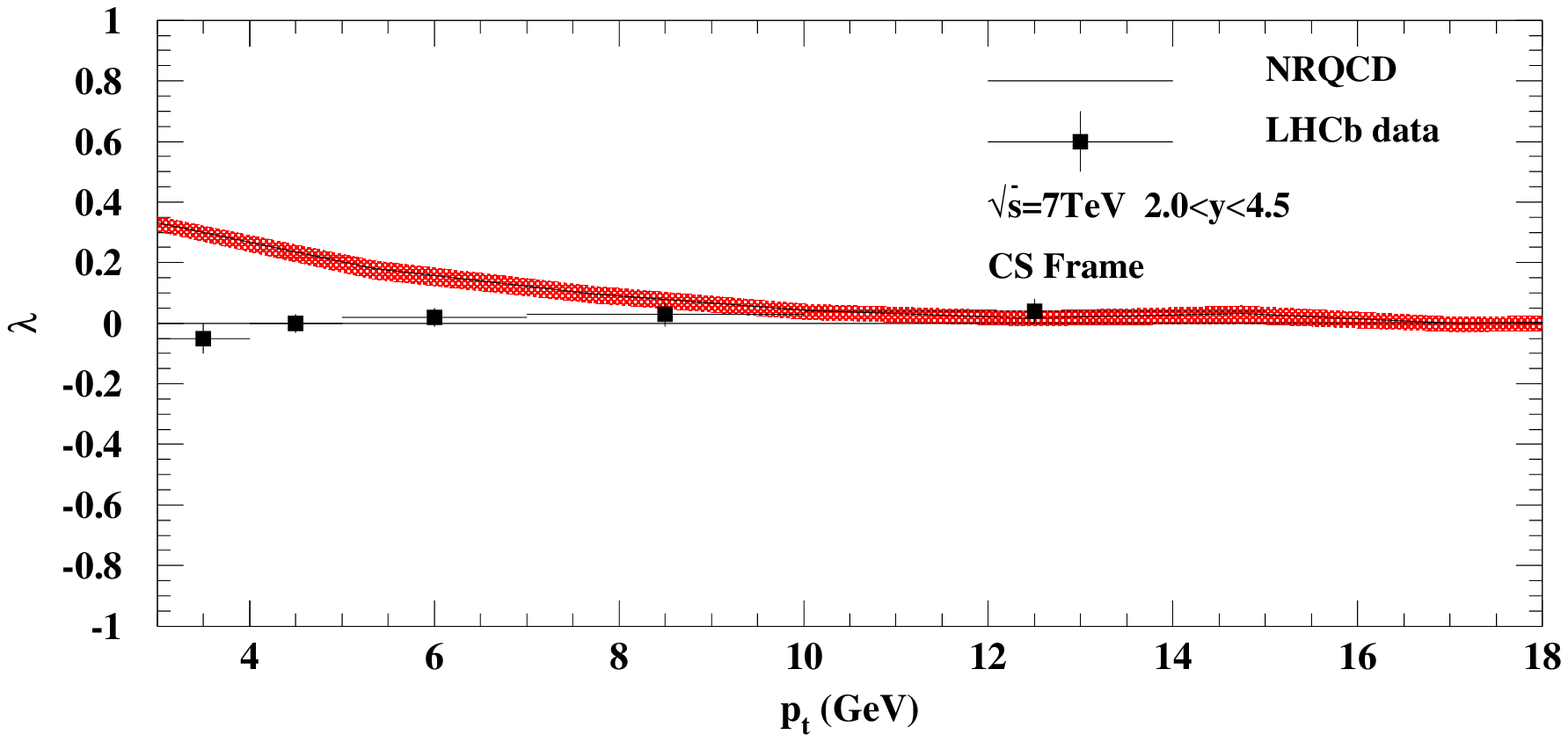}\\
\includegraphics*[scale=0.4]{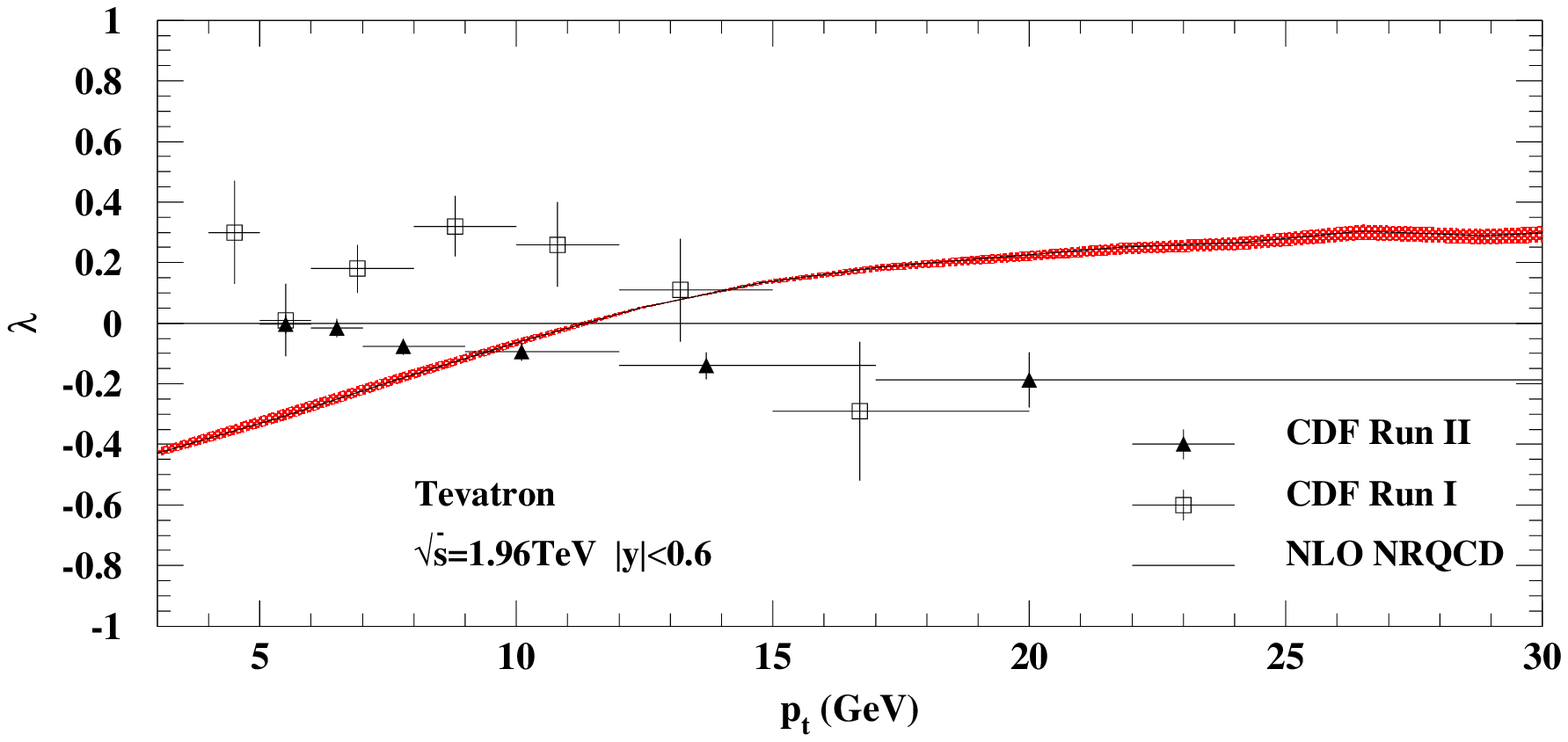}
\caption {\label{fig:pol}
$J/\psi$ polarization parameter $\lambda$ as a function of $p_t$.
The CDF and LHCb data are taken from Ref.~\cite{Affolder:2000nn, Abulencia:2007us, Aaij:2013nlm}
}}
\end{figure}

In Fig.\ref{fig:pol}, we present the results for $J/\psi$ polarization and compare them with the LHCb~\cite{Aaij:2013nlm} and CDF~\cite{Affolder:2000nn, Abulencia:2007us} data.
Our predictions can reproduce the LHCb data in both the helicity and the Collins-Soper frame at $p_t>7\gev$,
below which, perturbative calculations are believed not able to give reliable predictions.
The polarization curve for the Tevatron experimental condition pass through the two sets of CDF measurements in medium $p_t$ region.
Adopting the same treatment of the uncertainties as that of the yield ones,
the large errors in Eq.(\ref{eqn:LDMEJ}) finally result in very narrow bands of the $J/\psi$ polarization parameter,
which is actually easy to understand:
$M_0$ corresponds to the magnitude of contributions proportional to the $^1S_0^{[8]}$ one~\cite{Ma:2010jj},
even though its uncertainty is large, it contributes small and does not affect the results for $J/\psi$ polarization very much.

In summary, with the recent LHCb data of $\eta_c$ production rate, along with HQSS,
we obtained the CS LDMEs for both $J/\psi$ and $\eta_c$ directly by the fit of the experiment for the first time,
which is based on a thorough analysis of the uncertainties.
Our results are comparable with the values obtained in most of other existing works.
Using the relations of the LDMEs for $J/\psi$ production in Ref.~\cite{Shao:2014yta},
we also obtained the CO LDMEs for both $\eta_c$ and $J/\psi$ production,
which are consistent with the VSR.
Employing these LDMEs, our predictions on $\eta_c$ and $J/\psi$ hadroproduction rates are in good agreement with the CDF and LHCb data.
We also calculated the polarization of prompt $J/\psi$ at hadron colliders.
Our predictions can explain the LHCb data for $J/\psi$ polarization in both helicity and Collins-Soper frames,
and pass through the two sets of CDF measurements in medium $p_t$ region.
Our work provides another example to support NRQCD and an evidence for the HQSS and VSR.
It also helps to clarify the ambiguity of the determination of the CO LDMEs for $J/\psi$ production,
at the same time, opens a door to the solution to the long-standing $J/\psi$ polarization puzzel.

We are greatful to Jian-Xiong Wang for his generous help with the computer code.
We thank Sergey Barsuk, Maksym Teklishyn and Emi Kou for providing us with the experimental data on $\eta_c$ production.
We also thank Xing-Gang Wu, Bin Gong, Yang Ma, Hong-Hao Ma, Yu-Jie Zhang and Guang-Zhi Xu for helpful discussions.
This work is supported by the National Natural Science Foundation of China (Nos.~11405268 and Nos.~11105152),
and by the Fundamental Research Funds for the Central Universities under Grant No. SWU114003.

{\it Note added}-When our calculation was finished and the manuscript was preparing for publication,
we noticed two independent preprints~\cite{Butenschoen:2014dra, Han:2014jya} about the same topic.
However, our work contains something new and interesting.
On one hand, we noticed that $\eta_c$ hadroproduction process provided an excellent opportunity to fix the CS LDMEs,
and through a thorough analysis of the uncertainties, we obtained a reasonable range of these LDMEs.
On the other hand, our work not only supported CO mechanism, but also suggested an evidence for the HQSS and VSR.
The CO LDMEs obtained in our work are consistent with the VSR,
and are able to explain the $J/\psi$ and $\eta_c$ production experiment as well as the LHCb data on $J/\psi$ polarization in good precision.
Besides, the $p_t$ behavior of our prediciton on $\eta_c$ hadroproduction is consistent with the experiment.


\end{document}